\documentclass[aps,prl,twocolumn,groupedaddress,showkeys,showpacs]{revtex4}

\usepackage{graphicx}

\begin{document}

\title{Local fluctuations in the aging of a simple glass}

\author{Horacio E. Castillo$^{*}$ and Azita Parsaeian}
\affiliation{Department of Physics and Astronomy, Ohio University,
  Athens, OH, 45701, USA}

\date{May 11, 2006}

\pacs{PACS: 64.70.Pf, 61.20.Lc, 61.43.Fs}
\keywords{glass-forming liquids, spatially heterogeneous dynamics,
  relaxation, aging, nonequilibrium dynamics, Lennard-Jones mixture,
  supercooled liquid, molecular-dynamics} 

\maketitle

  {\bf 
  The presence of {\em dynamical heterogeneities},
  i.e. nanometer-scale regions containing molecules rearranging
  cooperatively at very different rates compared to the
  bulk~\cite{Ediger_review00,Sillescu_review99}, is increasingly being
  recognized as crucial in our understanding of the glass transition,
  from the non-exponential nature of relaxation, to the divergence of the relaxation times~\cite{AdamGibbs}. 
  Recently, dynamical heterogeneities have been directly
  observed
  experimentally~\cite{Kegel-Blaaderen-science00,Weeks-Weitz,Courtland-Weeks-jphysc03,Israeloff-AFM}
  and in simulations~\cite{Glotzer_simulations}. However a clear
  physical picture for the origin of these heterogeneities is still
  lacking. Here we investigate a possible physical mechanism for the
  origin of dynamical heterogeneities in the non-equilibrium dynamics
  of structural glasses. We test the predictions regarding universal
  scaling of fluctuations derived from this mechanism against
  simulation results in a simple binary Lennard-Jones glass model, and
  find that to a first approximation they are satisfied. We also
  propose to apply the same kind of analysis to experimental data from
  confocal microscopy in colloidal glasses.
}

Supercooled liquids approaching the glass transition display
increasingly slow dynamics, until eventually they cannot equilibrate
in laboratory timescales~\cite{Zallen83}. 
One consequence of this fact 
is {\em physical
aging}, i.e. the breakdown of {\em time translation invariance} (TTI):
the correlation $C(t,t_w)$ between
spontaneous fluctuations of an observable at times $t$ (the final
time) and $t_w$ (the waiting time) are nontrivial functions of $t$ and
$t_w$, as opposed to being functions of the time difference $t-t_w$. 
In many cases, the two-time correlation $C(t,t_w)$ in an aging system
separates into a fast, time translation invariant contribution
$C_{\mbox{\scriptsize fast}}(t-t_w)$, and a slow contribution
$C_{\mbox{\scriptsize slow}}(t,t_w)$~\cite{bckm_review97}: $ C(t,t_w)
= C_{\mbox{\scriptsize fast}}(t-t_w) + C_{\mbox{\scriptsize
slow}}(t,t_w)$. In the case of a structural glass, two-step relaxation 
is observed: the fast
term corresponds to localized fluctuations of individual particles
inside their cages, and the slow term corresponds to longer time
scales, in which cages break down and the system structurally relaxes.
For some systems, 
the slow part of the correlation has the
form~\cite{bckm_review97} $ C_{\mbox{\scriptsize slow}}(t,t_w) =
C_{\mbox{\scriptsize slow}}(h(t)/h(t_w))$, where $h(t)$ is some
monotonically increasing function. For example, in the case of 
domain growth, $h(t)$
is proportional to the domain size~\cite{bckm_review97}. 

Recently, 
it has been proven that, in the
limit of long times, the dynamics of a class of spin-glass models is invariant
under {\em global} reparametrizations $t \to h(t)$ of the time~\cite{ckcc-rpg-prl02}. This
result has been used to predict the existence of a Goldstone mode in
the nonequilibrium dynamics, 
associated with smoothly varying local fluctuations in the
reparametrization of the time $t \to h_r(t) = {\rm e}^{\varphi_r(t)}$.
These fluctuations have been physically interpreted to represent local
fluctuations of the age of the sample~\cite{ccck-prl-numeric,cccki-prb}. 
In the cases where 
the {\em global} two-time correlation exhibits
$h(t)/h(t_w)$ scaling, a simple Landau theory
approximation for the dynamical action
predicts~\cite{ccck-prl-numeric,cccki-prb,chamon-charb-cug-reich-sellito_condmat04} 
that the full probability
distribution $\rho(C_r(t,t_w))$ of {\em local} correlations $C_r(t,t_w)$ depends
only on the global correlation $C_{\mbox{\scriptsize global}}(t,t_w)$. 
In all of this discussion, only the slow
part of the correlation is considered, and any effects due to the fast
part of the dynamics are neglected.

In this Letter, we examine 
whether this theoretical
picture can be extended to explain dynamical heterogeneities in the aging of
structural glasses, by presenting the first detailed characterization
of the behavior of fluctuations in the aging of a continuous-space,
quasi-realistic structural glass model. The presence of a global
symmetry under time reparametrization in the generating functional for
the dynamics has been proven only for a class of spin glass models,
and it is far from obvious whether it applies in off-lattice,
quasi-realistic models, describing structural glasses.
Addtionally, the question about the aging behavior of local
observables in structural glasses remains mostly open. Almost all 
simulations in glass forming liquids have focused on the (equilibrium)
supercooled
liquid~\cite{Glotzer_simulations},
although there are some recent results for kinetically constrained
spin models of
glassyness~\cite{chamon-charb-cug-reich-sellito_condmat04}, and an
earlier work exploring dynamic spatial correlations for soft spheres in the
aging regime~\cite{parisi_jpcb99}.

We probe 
individual 
particle displacements along one direction $\Delta x_j(t, t_w) = x_j(t) -
x_j(t_w)$ (where $j$ is the particle index); and also {\em local,
coarse-grained} two-time functions: the correlator
\begin{equation}
C_{{\bf r}}(t, t_w) = \frac{1}{N(B_{\bf r})} \sum_{{\bf r}_j(t_w) \in B_{\bf r}} 
  \cos({\bf q} \cdot ({\bf r}_j(t) - {\bf r}_j(t_w))), 
\label{eq:Cqr_def}
\end{equation}
and the mean square displacement
\begin{equation}
\Delta_{\bf r}(t, t_w) = \frac{1}{N(B_{\bf r})} \sum_{{\bf r}_j(t_w) \in B_{\bf r}} 
  ({\bf r}_j(t) - {\bf r}_j(t_w))^2.
\label{eq:d2r_def}
\end{equation}
Here we consider a coarse graining cubic shaped box $B_{\bf r}$
of side $\ell$ around the point ${\bf r}$ in the
system, and the sums run over the $N(B_{\bf r})$ particles
{\em present at the waiting time $t_w$} in the box $B_{\bf r}$. 
We choose a value of q that corresponds to the main
peak in the structure factor $S(q)$ of the system, $q = 7.2$ in Lennard-Jones
units. 

These definitions are inspired by the analogous 
definitions in the case of spin
glasses~\cite{ccck-prl-numeric,cccki-prb}, and {\em can be applied both to
analyze data obtained from simulations and from confocal
microscopy experiments}. 
The {\em global} quantities $C_{\mbox{\scriptsize global}}(t, t_w)$
(incoherent part of the intermediate scattering function) and
$\Delta_{\mbox{\scriptsize global}}(t, t_w)$ (mean square
displacement) are defined by extending the sum to the whole system in
Eq.~(\ref{eq:Cqr_def}) and Eq.~(\ref{eq:d2r_def}) respectively. 

We performed 250 independent molecular dynamics runs for the binary
Lennard-Jones (LJ) system of Ref.~\cite{Kob-Barrat-aging-prl97}, which
has a mode coupling critical temperature $T_c =0.435$. 
A system of 8000 particles was
equilibrated at a temperature $T_0 = 5.0$, then instantly quenched 
to $T = 0.4$,
and finally it was allowed to evolve for $10^5$ Lennard-Jones time units. 
The origin of times was taken at the instant of the quench. 

\begin{figure}[ht]
\begin{center}

  \includegraphics[width=3.5in]{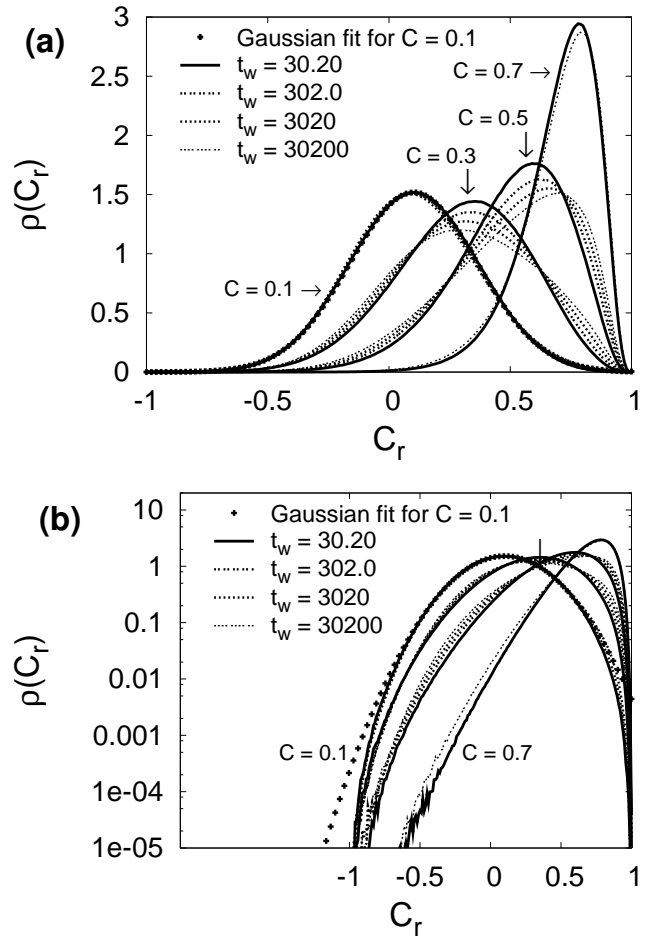}

\end{center}
\caption{ Probability distributions $\rho(C_r(t,t_w))$ for values
  $t_w$ in the range 30.20 -- 30200 (as indicated by the key in the
  figure), plotted for final times $t$ chosen so that
  $C_{\mbox{\scriptsize global}}(t, t_w) \in \{ 0.1, 0.3, 0.5, 0.7
  \}$.  Coarse graining box size $\ell \approx 0.11 L$, (with $L
  \equiv$ linear size of the simulation box). The curves collapse into
  four groups, corresponding to $C_{\mbox{\scriptsize global}} = 0.7, 0.5, 0.3,
  0.1$ (ordered from highest to lowest value of $C_r$ at the peak). A
  gaussian fit to the data for $C = 0.1$ is also shown. Top panel:
  linear scale. Bottom panel: logarithmic scale. }
\label{fig:rho-C}
\end{figure}

\begin{figure}[ht]
\begin{center}

  \includegraphics[width=3.5in]{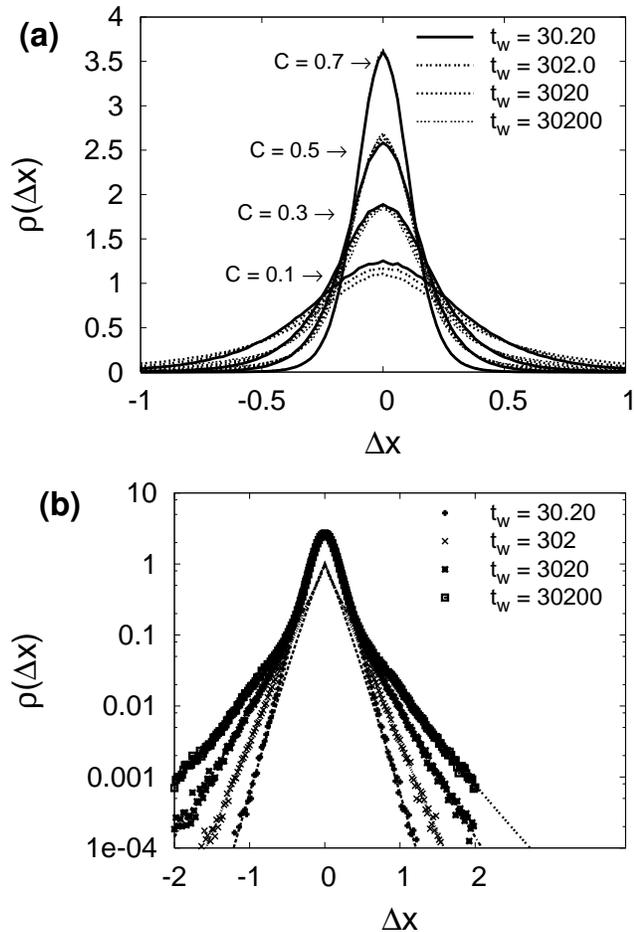}

\end{center}
\caption{ Top panel: Probability distributions $\rho(\Delta x(t,t_w))$
  for values $t_w$ in the range 30.20 -- 30200 (as indicated by the
  key in the figure), plotted for final times $t$ chosen so that
  $C_{\mbox{\scriptsize global}}(t, t_w) \in \{ 0.1, 0.3, 0.5, 0.7
  \}$.  
  The curves collapse into four groups, corresponding to
  $C_{\mbox{\scriptsize global}} = 0.1, 0.3, 0.5, 0.7$ (ordered from lowest to
  highest value of $\rho(\Delta x)$ at the peak). Bottom panel: Tails
  of the distributions $\rho(\Delta x(t,t_w))$, for $C = 0.5$ and $t_w
  = 30.20, 302, 3020, 30200$ (from narrower to wider tail). 
  Symbols: results from simulation. Lines:
  fits to the data for $|\Delta x| > 0.5$ using a stretched
  exponential form $\rho(\Delta x(t,t_w)) \approx {\cal N}
  \exp(-|\Delta x/a|^{\beta})$, with exponents $\beta = 1.11, 1.01,
  0.90, 0.81 $ respectively.}
\label{fig:rho-dx}
\end{figure}

\begin{figure}[ht]
\begin{center}

  \includegraphics[width=3.5in]{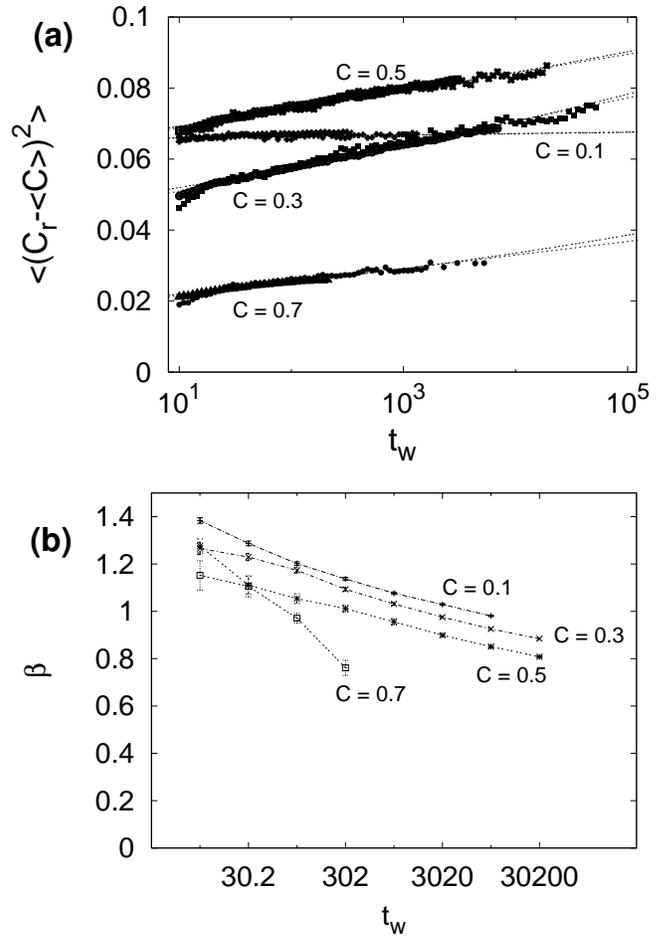}

\end{center}
\caption{ Evolution of the probability distributions, as a function of
  the waiting time $t_w$, at constant $C_{\mbox{\scriptsize global}}(t, t_w) \in
  \{ 0.1, 0.3, 0.5, 0.7\}$. Top panel: second moment of
  $\rho(C_r(t,t_w))$, together with fits to the functional forms: $m_0
  (t_w)^a$ (full lines) and $m_0^{'} \log(t_w/t_0)$ (dotted lines).
  Bottom panel: stretching exponent $\beta$ for the tails of
  $\rho(\Delta x)$, as a function of
  the waiting time $t_w$, at constant $C_{\mbox{\scriptsize global}}(t, t_w) \in
  \{ 0.1, 0.3, 0.5, 0.7\}$ (the lines are only guides to the eye). }
\label{fig:moment-drift}
\end{figure}

As mentioned above, we want to test whether the probability
distributions of local correlations depend on the two times $t$, $t_w$
only through the values of a global correlation. Unlike in spin
systems, in the case of a structural glass there are different global
correlation functions associated with the microscopic degrees of
freedom in the system, including the self-intermediate scattering function
$C_{\mbox{\scriptsize global}}(t, t_w)$ and the mean square displacement
$\Delta_{\mbox{\scriptsize global}}(t, t_w)$. 
This
leads us to testing two possible hypothesis: a) that the probability
distributions coincide when $C_{\mbox{\scriptsize global}}(t, t_w)$ is
kept constant; and b) that the probability distributions coincide when
$\Delta_{\mbox{\scriptsize global}}(t,t_w)$ is kept
constant. Extensive checking of our numerical data indicates that a)
is approximately satisfied by the probability distributions of 
$C_r(t,t_w)$, $\Delta_{\bf r}(t, t_w)$ and $\Delta x(t, t_w)$, 
but b) is satisfied only to a much lesser
degree~\cite{castillo-parsa_long-rhoC}.

In Fig.~\ref{fig:rho-C} we present our results for the probability
distribution $\rho(C_r(t,t_w))$ of the local intermediate scattering
function for waiting times $t_w = 30.20, \cdots, 30200$, 
and final times $t$ chosen so that 
$C_{\mbox{\scriptsize global}}(t, t_w) \in \{ 0.1, 0.3,
0.5, 0.7 \}$. We observe that the data approximately collapse for each
value of $C_{\mbox{\scriptsize global}}(t, t_w)$. This kind of
collapse is also observed in simulations in a 3D spin glass model, but
in the case of the spin glass model, the collapse is more precise than
here. Unlike the case of the 3D spin glass model, the position of the
peak in the distribution $\rho(C_r)$ is strongly dependent on the
value of $C_{\mbox{\scriptsize global}}(t, t_w)$. The skewness of
$\rho(C_r)$ also depends dramatically on $C_{\mbox{\scriptsize
global}}(t, t_w)$: it goes from highly skewed for $C_{\mbox{\scriptsize global}}(t,t_w) =
0.7 $ to almost symmetric around its peak for $C_{\mbox{\scriptsize global}}(t,t_w) =
0.1$. As shown in the figure, the distributions $\rho(C_r(t,t_w))$ for
$C_{\mbox{\scriptsize global}}(t,t_w) = 0.1$ can be well approximated by a Gaussian fit,
but this does not hold for larger values of $C_{\mbox{\scriptsize global}}(t,t_w)$. 
Unlike previous results in kinetically constrained
models~\cite{chamon-charb-cug-reich-sellito_condmat04}, in the present
case, Gumbel distributions {\em do not } provide good fits to $\rho(C_r)$.
To characterize the weak dependence of the probability distributions
on waiting time at fixed $C_{\mbox{\scriptsize global}}(t, t_w)$, in
the top panel of Fig.~\ref{fig:moment-drift} we plot the centered second moment of the
distributions $\rho(C_r)$ as a function of
waiting time, for fixed $C_{\mbox{\scriptsize global}}(t, t_w) \in \{
0.1, 0.3, 0.5, 0.7 \}$. The dependence on
$t_w$ is so weak that both a logarithmic form and a power law form
(with powers in the range $0.01-0.07$) provide a good fit. 

In Fig.~\ref{fig:rho-dx} we present our results for the probability
distribution $\rho(\Delta x(t,t_w))$ of the particle displacements
$\Delta x_j(t,t_w) = x_j(t)-x_j(t_w)$ along one direction. In the top
panel of Fig.~\ref{fig:rho-dx} we can observe that these data also
approximately collapse for each value of $C_{\mbox{\scriptsize
global}}(t, t_w)$. In the bottom panel we have a closer look at the
tails of $\rho(\Delta x(t,t_w))$. We find that the distribution is
non-gaussian, as was observed in experiments in colloidal glasses in
the supercooled regime~\cite{Weeks-Weitz}. The tails of the
distribution can be fitted with a stretched exponential form
$\rho(\Delta x) \approx {\cal N} \exp(-|\Delta x/a|^{\beta}) $, and
they become more prominent as $t_w$ grows (for  constant
$C_{\mbox{\scriptsize global}}(t, t_w)$). Indeed, as shown in the bottom panel
of Fig.~\ref{fig:moment-drift}, the exponent $\beta$ decreases from
values above unity at short waiting times $t_w$ to values of around $0.8$ at
much longer $t_w$.

To summarize, in this Letter we have presented the first detailed
characterization of non-equilibrium fluctuations in the aging regime
in a continuous-space, quasi-realistic structural glass model. It is
also the first test for the possible presence in such models of a
Goldstone mode associated with reparametrizations of the time
variable, which could explain the behavior of fluctuations in the
aging regime. 
We have found that, as expected from the Goldstone mode picture,
the probability distributions for the local fluctuating two-time
quantities are, to a first approximation, invariant when the global
intermediate scattering function $C_{\mbox{\scriptsize global}}(t,
t_w)$ is kept constant. 

Besides the scaling predicted by the Goldstone mode picture, a
detailed analysis of the probability distributions for local
observables uncovers that: i)
unlike the case of spin glasses, the position of the peak for the
probability distributions of the local correlation $\rho(C_r(t,t_w))$
shifts dramatically as a function of the value of
$C_{\mbox{\scriptsize global}}(t, t_w)$; ii) the distribution
$\rho(C_r(t,t_w))$ evolves gradually from being highly skewed and
non-gaussian for larger values of $C_{\mbox{\scriptsize global}}(t,
t_w)$ to being unskewed and very close to gaussian for small values of
$C_{\mbox{\scriptsize global}}(t, t_w)$; iii) a more detailed
examination of the probability distributions at constant
$C_{\mbox{\scriptsize global}}(t, t_w)$ reveals that the moments of
the distributions evolve smoothly as a function of the waiting time
$t_w$, without displaying any obvious characteristic timescale (this
may be due in part to the fact that the dynamic correlation length in
the system is growing as a function of $t_w$, but for the timescales
of the simulation it is not yet larger than the size of the coarse
graining box used~\cite{castillo-parsa_lengthscale}, leading to a
gradual increase in the variance of all coarse grained quantities) ;
iv) the probability distributions of one-dimensional displacements
$\rho(\Delta x (t,t_w))$ are clearly non-gaussian, as in confocal
microscopy experiments {\em in the supercooled liquid}
regime~\cite{Weeks-Weitz}, and their tails can be fit by stretched
exponential forms, with exponents $\beta$ that decrease from $\beta >
1$ to $\beta \approx 0.8$ as $t_w$ increases (for fixed
$C_{\mbox{\scriptsize global}}(t, t_w)$).

We conclude that the probability distributions of local observables in
the aging regime are consistent with the presence of a Goldstone mode
controlling the nonequilibrium dynamics of a structural glass. We
suggest that a direct experimental test of this picture could be
provided by applying a similar analysis to experimental data from
confocal microscopy in colloidal glass systems.

H.E.C. especially thanks C.~Chamon and L.~Cugliandolo for very enlightening
discussions over the years, and J.~P.~Bouchaud, D.~Reichman, and
E.~Weeks for suggestions and discussions. This work was supported in
part by DOE under grant DE-FG02-06ER46300, by NSF under grant PHY99-07949,
and by Ohio University. 
Numerical simulations were carried out at the Ohio Supercomputing
Center and at the Boston University SCV. 
H.E.C. acknowledges the hospitality of the Aspen
Center for Physics.

\end{document}